\documentclass[sigconf,natbib=true]{acmart}
\usepackage{pbalance}
\usepackage{balance}
\usepackage{enumitem}
\usepackage{booktabs}
\usepackage{graphicx}
\usepackage{array}
\usepackage{lineno}
\usepackage{multirow}
\usepackage{hhline}
\usepackage{makecell}
\usepackage{subfigure}
\usepackage[marginal]{footmisc}

\AtBeginDocument{%
  \providecommand\BibTeX{{%
    \normalfont B\kern-0.5em{\scshape i\kern-0.25em b}\kern-0.8em\TeX}}}

\copyrightyear{2025}
\acmYear{2025}

\begin{document}

\title{
MBGR: Multi-Business Prediction for Generative Recommendation at Meituan
}


\author{Changhao Li}
\authornote{Both authors contributed equally to this research.}

\affiliation{%
  \institution{Meituan}
  \city{Chengdu}
  \country{China}
}
\email{lichanghao@meituan.com}

\author{Junwei Yin}
\authornotemark[1]

\affiliation{%
  \institution{Meituan}
  \city{Chengdu}
  \country{China}
}
\email{yinjunwei03@meituan.com}

\author{Zhilin Zeng}
\affiliation{%
\institution{Meituan}
   \city{Chengdu}
  \country{China}
  }
\email{zengzhilin@meituan.com}

\author{Senjie Kou}
\affiliation{%
\institution{Meituan}
   \city{Chengdu}
  \country{China}
  }
\email{kousenjie@meituan.com}

\author{Shuli Wang}
\authornote{Corresponding author.}
\affiliation{%
  \institution{Meituan}
   \city{Chengdu}
  \country{China}
}
\email{wangshuli03@meituan.com}

\author{Wenshuai Chen}
\affiliation{%
\institution{Meituan}
   \city{Chengdu}
  \country{China}
  }
\email{chenwenshuai@meituan.com}

\author{Yinhua Zhu}
\affiliation{%
\institution{Meituan}
   \city{Chengdu}
  \country{China}
  }
\email{zhuyinhua@meituan.com}

\author{Haitao Wang}
\affiliation{%
\institution{Meituan}
   \city{Chengdu}
  \country{China}
  }
\email{wanghaitao13@meituan.com}

\author{Xingxing Wang}
\affiliation{%
\institution{Meituan}
   \city{Beijing}
  \country{China}
  }
\email{wangxingxing04@meituan.com}

\renewcommand{\shortauthors}{Changhao Li et al.}



\begin{abstract}
Generative recommendation (GR) has recently emerged as a promising paradigm for industrial recommendations. GR leverages Semantic IDs (SIDs) to reduce the encoding-decoding space and employs the Next Token Prediction (NTP) framework to explore scaling laws. However, existing GR methods suffer from two critical issues: (1) a \textbf{seesaw phenomenon} in multi-business scenarios arises due to NTP’s inability to capture complex cross-business behavioral patterns; and (2) a unified SID space causes \textbf{representation confusion} by failing to distinguish distinct semantic information across businesses.  
To address these issues, we propose Multi-Business Generative Recommendation (MBGR), the first GR framework tailored for multi-business scenarios. Our framework comprises three key components.  
First, we design a Business-aware semantic ID (BID) module that preserves semantic integrity via domain-aware tokenization. Then, we introduce a Multi-Business Prediction (MBP) structure to provide business-specific prediction capabilities. Furthermore, we develop a Label Dynamic Routing (LDR) module that transforms sparse multi-business labels into dense labels to further enhance the multi-business generation capability. 
Extensive offline and online experiments on Meituan’s food delivery platform validate MBGR’s effectiveness, and we have successfully deployed it in production.  

\end{abstract}

\ccsdesc[500]{Information systems~Retrieval models and ranking}
\ccsdesc[500]{Information systems~Computational advertising}

\keywords{Recommender Systems, Generative recommendation, Multi-business}



\maketitle

\section{Introduction}
Internet technology has made online businesses like e-commerce platforms and social networks essential to daily life \citeN{W&D, deepfm, xdeepfm}. Today's major platforms provide diverse businesses - Meituan, for instance, spans food delivery, entertainment and healthcare. While multi-business modeling captures shared patterns, business heterogeneity and varying user preferences pose challenges for personalized recommendations \citeN{din, dien, sim}.

The core challenge of multi-business modeling is balancing business-specific characteristics while learning shared representations across businesses \citeN{li2020HMOE, chang2023pepnet}. Traditional approaches typically employ multiple expert towers \citeN{zhou2023hinet, ple, sheng2021star}to model both business commonalities and unique features, with techniques like causal inference and LLM-enhanced fusion showing promising results. However, conventional recommendation systems face two persistent limitations: substantial storage requirements from item embeddings and poor long-tail performance. The emerging generative recommendation paradigm addresses these by using compact semantic IDs to represent items, achieving both storage reduction and direct target generation. Specifically, it predicts users' next interactions by generating semantic IDs from historical sequences, demonstrating superior training efficiency.

However, current generative recommendation frameworks \citeN{tiger, hstu, onerec, mtgr} are designed for a single business and lack multi-business optimization. This results in separate deployments for each business, increasing model training and maintenance costs. Furthermore, it is difficult to leverage historical user data from other businesses to enhance the model. This raises a key question: how can we effectively integrate different businesses and achieve efficient generative recommendations? Designing a multi-business-specific generative recommendation framework is necessary, but it faces two key challenges: First, the signals from multiple businesses are mixed and difficult to distinguish. In real-world scenarios, various businesses often influence each other, creating complex relationships. User behavior across different businesses can vary significantly, leading to significant heterogeneity between businesses. Furthermore, the sheer volume of data across businesses makes it difficult to extract commonalities and differences from the mixed signals. Second, multiple businesses share a common semantic codebook, making optimization difficult. Compressing heterogeneous data from multiple businesses onto semantic IDs in the same semantic space results in gradient coupling during optimization, making it impossible to account for inter-business differences when generating semantic IDs for items, resulting in poor performance. These two challenges hinder the application of generative recommendations in multi-business scenarios and limit recommendation effectiveness, necessitating innovative solutions.

To address the above challenges, we propose MBGR, a Generative Recommendation framework specifically designed for Multi-business Prediction. MBGR consists of three core modules: the Business-aware semantic ID (BID) module, the Multi-Business Prediction (MBP) module, and the Label Dynamic Routing (LDR) module.
The BID module addresses the gradient coupling problem caused by semantic IDs in the same semantic space by designing an encoder and decoder. It innovatively introduces a business encoding representing the business data context on top of the original semantic ID to achieve business adaptation. The encoder obtains business-related item representations, and the decoder reconstructs the semantic ID to minimize semantic information loss. 
The MBP module introduces a parameter-sharing Mixed Tower of Experts (MoE) architecture, performing weighted attention aggregation on each expert tower to obtain the final business-related representation. 
The LDR module searches for the label of the next interaction with the same business from the user interaction sequence, which transforms sparse multi-business labels into dense labels to further enhance the multi-business generation capability. 


Our contributions can be summarized in three points:

\begin{itemize}
\item Multi-business Generative Recommendation Framework: We propose the MBGR framework, which effectively combines generative and multi-business recommendation, significantly reducing online item storage and model maintenance costs, and providing a new approach to multi-business learning. To our knowledge, this is the first work to combine generative and multi-business recommendation.


\item Multi-business Signal Separation: We proposed a collaborative mechanism between the SID, MBP, and LDR modules to effectively separate the mixed multi-business signals from two dimensions: model structure (hybrid tower of experts) and label routing. This ensures knowledge sharing between businesses while maintaining the specificity of each business, providing a complete solution for generative recommendation in multi-business scenarios.

\item Extensive Experimental Validation: We conducted extensive offline and online experiments and achieved a 3.98\% improvement in CTCVR online, demonstrating the effectiveness of our proposed MBGR.
\end{itemize}

\section{Related Work}

\subsection{Generative Recommendations}
Generative recommender systems aim to replace the traditional two-tower recall model by directly generating candidate items. They have garnered widespread attention in the industry in recent years. Early work primarily explored representing items as sequences of semantic IDs. TIGER \cite{tiger} first proposed using RQ-VAE \cite{rqvae} residual quantization encoding to convert items into semantic IDs and using Transformer parameters as an index to directly generate candidate sets, effectively alleviating sparsity and cold-start issues. HSTU \cite{hstu} further extended generative ideas to the ranking stage, unifying heterogeneous feature spaces and redefining recall ranking as a generative task. This was the first demonstration of the applicability of scaling laws in recommender systems. Subsequent work has focused on optimizing efficiency. OneRec \cite{onerec} proposed an end-to-end architecture to replace the traditional cascade model. It improves tokenization by introducing collaborative prior knowledge and employs an encoder-decoder structure to enhance computing power utilization. COBRA \cite{cobra} proposed a cascade representation method to address the information loss problem, combining sparse and dense representations to process semantic IDs and fine-grained information respectively. In terms of application expansion, LC-Rec \cite{lcrec} explored the integration of LLM and recommendation systems, allowing LLM to learn item collaborative semantics through instruction fine-tuning; GenRank \cite{GenRank} systematically studied the influencing factors of generative ranking and verified the importance of sequence interaction methods and feature engineering. To address the issue of inference efficiency, RPG \cite{rpg} proposed a parallel semantic ID generation framework, using product quantization to support the unordered parallel generation of long semantic IDs. The latest work has begun to focus on commercial applications. EGA-V2 \cite{ega-v2} extends the generative architecture to the advertising system, uniformly handling multiple links such as bidding and creative selection; PinRec \cite{pinrec} proposes a result-oriented generation method, meeting the needs of different business goals through conditional generation.

\subsection{Multi-business recommendations}
Multi-business (scenario) recommendation systems aim to leverage data from multiple scenarios to train a unified model to improve recommendation effectiveness. Early research, starting from a causal perspective, first proposed a causal-inspired intervention framework, CausalInt \cite{causalint}, addressing the multi-scenario modeling problem through invariant representation modeling and an inter-scenario transfer module. M-scan \cite{mscan} further discovered that the direct influence of context on click behavior can lead to prediction bias, and proposed a context-aware collaborative attention mechanism and a context bias eliminator to alleviate this problem. In terms of architectural design, M3OE \cite{m3oe} designed three hybrid expert modules to learn shared, context-specific, and task-specific user preferences, respectively, and proposed a two-level fusion mechanism for precise control. SESRec \cite{sesrec} specifically targets joint search and recommendation modeling, designing a context-specific interest extraction layer and a global label space multi-task layer to effectively fuse search and recommendation information. Considering the needs of industrial applications, IncMSR \cite{incmsr} proposed an incremental learning method that improves training efficiency and senses distribution changes by quantifying representation distances across different dimensions and applying fine-grained constraints. Recent research has begun exploring the power of large models. LLM4MSR \cite{llm4msr} leverages large language models to capture cross-scenario relevance and personalized preferences, connecting the semantic space and the recommendation space through a hierarchical meta-network.

\section{Problem Formulation}

In this section, we formally define the multi-business generative sequential recommendation problem. Given a user $u$'s historical interaction sequence across multiple businesses:
\begin{equation} \label{equation:1}
S_u = \{s_{u,1}, s_{u,2}, ..., s_{u,L}\}
\end{equation}

where each interaction $s_{u,t} = (i_{u,t}, b_{u,t}, t_{u,t})$ consists of:
\begin{itemize}
    \item $i_{u,t}$: item ID
    \item $b_{u,t}$: business type
    \item $t_{u,t}$: timestamp
\end{itemize}

Our goal is to simultaneously generate the Semantic ID of the next possible item for each of the $K$ business types $b_k \in B$ ($k=1,...,K$) using a generative approach.

Formally, we aim to maximize the following conditional probability:
\begin{equation}
P(T_u^{(1:K)}|S_u) = \sum_{k=1}^K P(T_u^{(k)}|S_u, b_k)
\end{equation}

where $T_u^{(k)} = \{t_1^{(k)}, t_2^{(k)}, ..., t_{L_k}^{(k)}\}$ represents the Semantic ID token sequence of the next item in business $b_k$, and $L_k$ is the length of the Semantic ID for business $b_k$.

\section{Proposed Method}

We propose an innovative multi-business generative recommendation system architecture that consists of three core components: 1) Business-aware semantic ID (BID) module, 2)Multi-Business Prediction (MBP) module, and 3) Label Dynamic Routing (LDR) module. 

\begin{figure*}[htbp]
\begin{center}
\includegraphics[width=\textwidth]{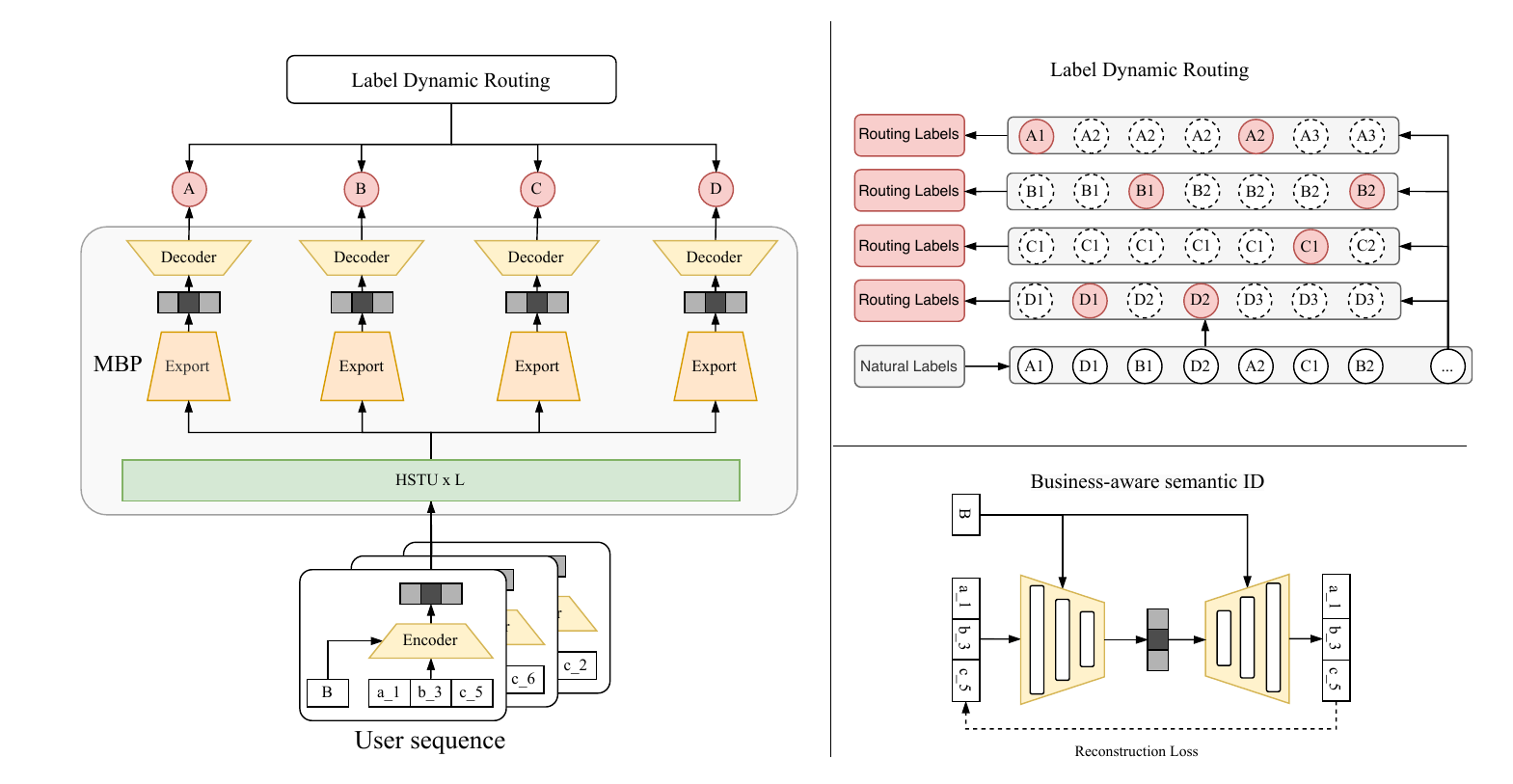}
\end{center}
\caption{The overall architecture of MBGR.}
\label{fig:mbgr}
\end{figure*}

\subsection{Business-aware semantic ID (BID)}

The Business-aware semantic ID (BID) module is a novel dual-path representation learning framework designed to address two critical challenges in multi-business recommendation: 1) incorporating business context into shared token representations, and 2) minimizing semantic information loss during the encoding process. The module employs a business-aware autoencoder architecture that serves two complementary purposes during training: input representation learning and next-item prediction.

The BID module consists of three core components:

1. \textbf{Business-Aware Encoder}:
The encoder processes input tokens to learn business-aware item representations, while the decoder reconstructs the original tokens to ensure semantic preservation.Transforms input token embeddings $\mathbf{t}_i = [\mathbf{t}_{i,1}, \mathbf{t}_{i,2}, ..., \mathbf{t}_{i,K}] \in \mathbb{R}^{K \cdot d_t}$ into business-aware item representations $\mathbf{e}_i \in \mathbb{R}^{d_e}$:

\begin{align}
    \mathbf{e}_i^{enc} &= \text{FFN}_{enc}([\mathbf{t}_i, \mathbf{b}_i]) \\
    \mathbf{g}_i^{enc} &= \sigma(\text{FFN}_{gate}^{enc}([\mathbf{e}_i^{enc}, \mathbf{b}_i])) \\
    \mathbf{e}_i &= \mathbf{e}_i^{enc} \odot \mathbf{g}_i^{enc}
\end{align}

2. \textbf{Business-Conditioned Decoder}:
Reconstructs token embeddings from item representations and generates next-item token sequences:

\begin{align}
    \hat{\mathbf{t}}_i^{dec} &= \text{FFN}_{dec}([\mathbf{e}_i, \mathbf{b}_i]) \label{eq:decoder1}\\
    \mathbf{g}_i^{dec} &= \text{ReLU}(\text{FFN}_{gate}^{dec}([\hat{\mathbf{t}}_i^{dec}, \mathbf{b}_i])) \label{eq:decoder2}\\
    \hat{\mathbf{t}}_i &= \hat{\mathbf{t}}_i^{dec} \odot \mathbf{g}_i^{dec}\label{eq:decoder3}
\end{align}

Key innovations of the BID module include:
\begin{itemize}
    \item \textbf{Dual-Path Architecture}: Simultaneously learns business-aware representations and predicts multi-business tokens
    \item \textbf{Semantic Preservation}: The reconstruction objective ensures minimal information loss during encoding
    \item \textbf{Business-Aware Encoding and Decoding}: Both the encoder and decoder incorporate business context through the business type embedding $\mathbf{b}_i$
    \item \textbf{Parameter Sharing}: The same decoder is used for both reconstruction and prediction, improving efficiency
\end{itemize}

This design enables the BID module to effectively balance the competing requirements of business-specific adaptation and semantic consistency

\subsection{Multi-Business Prediction (MBP)}
This architecture serves as the core generative engine that simultaneously predicts next items across multiple business domains. Built upon a Transformer-based autoregressive framework, the architecture processes user interaction sequences $S_u$ to generate Semantic ID token sequences for all business types in parallel.
 \subsubsection{Sequence Encoding}:
The historical interaction sequence $S_u$ is encoded by the BID-encoder module into item representations, where each event contains item ID $ i $, business type $ b $, and timestamp $ t $.

 \subsubsection{Business-Aware Item Representation}:
To bridge the gap between universal item representations and business-specific requirements, we propose a module that dynamically transforms shared embeddings into domain-specific representations. This module employs a Mixture-of-Experts (MoE) architecture with shared parameters, enabling efficient adaptation of generic item features to diverse business contexts.

Given a general item representation $\mathbf{e} \in \mathbb{R}^{d_e}$, the MBP module generates business-specific representations $\mathbf{e}^b$ for each business domain $b$ through a three-stage transformation process:

1. Contextual Fusion: Concatenate the item embedding with business context:
\begin{equation}
    \mathbf{z}^b = [\mathbf{e}, \mathbf{b}] \in \mathbb{R}^{d_e + d_b}
\end{equation}
where $\mathbf{b} \in \mathbb{R}^{d_b}$ is the learnable business type embedding that encodes domain-specific characteristics.

2. Adaptive Gating: Compute business-specific attention weights:
\begin{equation}
    \mathbf{g}^b = \text{SiLu}(\text{FFN}_{gate}(\mathbf{z}^b)) \in \mathbb{R}^K
\end{equation}
where $\text{FFN}_{gate}$ is a shared gating network that learns to allocate expert resources based on both item content and business context.

3. Expert Aggregation: Combine specialized transformations:
\begin{equation}
    \mathbf{e}^b = \sum_{k=1}^{K} g_k^b \cdot \text{FFN}_k^{exp}(\mathbf{z}^b) \in \mathbb{R}^{d_e}
\end{equation}
where $\text{FFN}_k^{exp}$ are $K$ shared expert networks that learn to extract business-relevant features from the fused representation $\mathbf{z}^b$.

\subsubsection{Business-Aware SID representation generation}:
The business-specific item representations $\mathbf{z}^b$ are decoded into token representations using the Business-Conditioned Decoder from BID.AS defined in Eqs.\eqref{eq:decoder1}-\eqref{eq:decoder3}\\
For each business type $b_k$, after obtaining the business-specific item representation $\mathbf{e}^b$, the decoder transforms it into Semantic ID token sequences:

\begin{equation}
    \hat{T}_u^{(k)} = \text{Decode}(\mathbf{e}^b, b_k)
\end{equation}

\subsection{Label Dynamic Routing}

For each position t in the sequence and each business type $b_k$, the prediction target is determined by finding the nearest future interaction of the same business type:
\begin{equation}
    i_{u,t+1}^{(k)} = i_{u,t'} \quad \text{where} \quad t' = \min\{t'' > t | b_{u,t''} = b_k\}
\end{equation}

During training, if no future interaction exists for a particular business type (i.e., $\nexists t' > t$ where $b_{u,t'} = b_k$), we mask the loss for that business to avoid invalid predictions. This approach ensures the model learns to predict meaningful next items for all active business types at each sequence position.

The architecture employs business-specific prediction heads that generate Semantic ID token sequences $T_u^{(k)} = \{t_1^{(k)}, t_2^{(k)}, ..., t_{L_k}^{(k)}\}$ for each business $b_k$ in an autoregressive manner.

The model is trained end-to-end using a multi-task objective that combines InfoNCE losses across all business domains, enabling simultaneous learning of business-specific patterns while maintaining shared representations.

\section{Model Training}

Our multi-business generative recommendation system is trained end-to-end using a carefully designed multi-objective loss function that combines contrastive learning with semantic reconstruction. The training process simultaneously optimizes for both accurate multi-business prediction and semantic consistency across all business domains.

\subsection{Training Objective}

The overall training objective combines InfoNCE loss for multi-business prediction with reconstruction loss for semantic preservation:

\begin{equation}
    \mathcal{L} = \mathcal{L}_{\text{infoNCE}} + \lambda \cdot \mathcal{L}_{\text{recon}}
\end{equation}

where $\lambda$ is a hyperparameter controlling the trade-off between prediction accuracy and semantic consistency.

\subsection{InfoNCE Loss for Multi-Business Prediction}

The InfoNCE loss is computed across all business domains and token positions to optimize the model's ability to predict the next item's Semantic ID tokens for different business:

\begin{equation}
    \mathcal{L}_{\text{infoNCE}} = -\sum_{b=1}^{B} w_b \cdot w_t \sum_{i=1}^{N} \sum_{k=1}^{K} \log \frac{\exp(\text{sim}(\hat{\mathbf{t}}_{i,k}^b, \mathbf{t}_{i,k}^b)/\tau)}{\sum_{j=1}^{|\mathcal{V}_k|} \exp(\text{sim}(\hat{\mathbf{t}}_{i,k}^b, \mathbf{v}_{j,k})/\tau)}
\end{equation}

where:
\begin{itemize}
    \item $B$ is the number of business domains
    \item $w_b$ is the weight for business $b$, balancing the contribution of different business types
    \item $w_t = \exp\left(-\alpha \cdot (t_{\text{last}} - t_{u,t+1}^{(k)})\right)$ is the time decay coefficient, where $\alpha$ is a positive hyperparameter controlling the decay rate, $t_{\text{last}}$ is the timestamp of the last interaction in the sequence, and $t_{u,t+1}^{(k)}$ is the timestamp of the target item
    \item $N$ is the number of items in the batch
    \item $K$ is the number of tokens per item's Semantic ID
    \item $\tau$ is the temperature parameter controlling the sharpness of the distribution
    \item $\hat{\mathbf{t}}_{i,k}^b$ is the predicted token embedding for the $k$-th token of item $i$ in business $b$
    \item $\mathbf{t}_{i,k}^b$ is the ground truth token embedding
    \item $\mathbf{v}_{j,k}$ represents negative samples from the vocabulary $\mathcal{V}_k$ for token position $k$
    \item $\text{sim}(\cdot,\cdot)$ computes the cosine similarity between embeddings
\end{itemize}

\subsection{Reconstruction Loss for Semantic Preservation}

To ensure minimal semantic information loss during the encoding process, we employ a reconstruction loss that measures the difference between original and reconstructed token embeddings:

\begin{equation}
    \mathcal{L}_{\text{recon}} = \frac{1}{K} \sum_{k=1}^K \| T_u^{(k)} - \hat{T}_u^{(k)} \|_2^2
\end{equation}

where $T_u^{(k)} = \{t_1^{(k)}, t_2^{(k)}, ..., t_{L_k}^{(k)}\}$ represents the original sequence of Semantic ID tokens and $\hat{T}_u^{(k)}$ represents the reconstructed sequence for business $b_k$.











\section{Experiments}\label{Experiments}
To comprehensively evaluate the effectiveness of our proposed MBGR framework, we conducted extensive experiments in both offline and online settings. This section presents our experimental methodology, including dataset descriptions, baseline methods, and evaluation protocols.

\subsection{Experimental Setup}
\subsubsection{Datasets}

We evaluate our framework using two distinct datasets corresponding to different phases of deployment:

1. \textbf{Generative Training Dataset}:
Collected from Meituan's platform, this dataset contains user interaction sequences spanning one year:
\begin{itemize}
    \item \textbf{User Base}: 38,258,649 unique users with click actions
    \item \textbf{Item Catalog}: 54,875,570 distinct merchants across four business categories
    \begin{itemize}
        \item Business A: 61.47\% of merchants
        \item Business B: 9.56\% of merchants
        \item Business C: 12.31\% of merchants
        \item Business D: 16.66\% of merchants
    \end{itemize}
    \item \textbf{Temporal Coverage}: One year of sequential user interactions
\end{itemize}

2. \textbf{Downstream Application Dataset}:
Used for the deployment phase, this larger-scale dataset features:
\begin{itemize}
    \item \textbf{User Base}: 37,349,276 unique users
    \item \textbf{Interaction Samples}: 783,946,360 user interaction records
    \item \textbf{Update Frequency}: Daily refresh to maintain the most recent 30-day window
\end{itemize}

\subsubsection{Baselines}
We compare our proposed MBGR framework against two state-of-the-art sequential recommendation models:

1. \textbf{SASRec} (Self-Attentive Sequential Recommendation):
SASRec employs a unidirectional Transformer architecture to capture sequential dependencies in user-item interactions for next-item prediction. The model utilizes self-attention mechanisms to weigh the importance of different historical interactions, enabling it to focus on the most relevant items when making predictions. SASRec serves as a strong baseline for sequential recommendation tasks, particularly in single-domain settings.

2. \textbf{HSTU} (Hierarchical Sequential Transduction Unit):
HSTU is a novel hierarchical Transformer architecture specifically designed for large-scale generative recommendation systems. The model introduces a multi-level transduction mechanism that can effectively handle high-cardinality item spaces and capture complex sequential patterns in user behavior.

\subsubsection{Evaluation Protocol}
We employ a two-part evaluation strategy that assesses both the generative training phase and online application phase:

1. \textbf{Generative Training Evaluation}:
For the generative training phase, we use Hit Rate@10 (HR@10) as the primary metric:
\begin{itemize}
    \item \textbf{Business-wise Analysis}: We compute HR@10 separately for each business category
    \item \textbf{Focus on Small Businesses}: Special attention is given to performance on smaller business categories (e.g., B, C) to evaluate the effectiveness of our business-aware learning approach
\end{itemize}

2. \textbf{Downstream Application Evaluation}:
For the deployment phase, we use Grouped Area Under Curve (GAUC) as the primary metric:
\begin{itemize}
    \item \textbf{Business-wise GAUC}: GAUC is calculated separately for each business category to assess business-specific performance.
\end{itemize}

The comprehensive evaluation demonstrates our framework's dual strengths in both generative capability (measured by HR@10) and downstream ranking performance (measured by GAUC), while maintaining robust system performance under production workloads.

\subsubsection{Implementation Details}
\textbf{Generative Training Phase}:
The generative model was implemented in PyTorch 2.7.0 using NVIDIA A100-80GB GPUs. We used Adam optimizer with learning rate 0.001 and initialized parameters with normal distribution ($\mu=0$, $\sigma=0.01$). The architecture featured 8 attention heads and 16 transformer layers with 128-dimensional embeddings, trained with batch size 1024. \textbf{The maximum sequence length was set to 1500 items, with truncation or padding applied to all input sequences.} Additional hyperparameters were tuned through grid search on the validation set.

\textbf{Downstream Application Phase}:
For the downstream application, we used TensorFlow 1.15 with identical model architecture, incorporating both user embeddings and BID-encoded item representations at the embedding layer. The model was trained on the same 30-day window and deployed on A100-80GB GPUs. The serving infrastructure included distributed embedding lookup and business-aware sharding to handle production traffic.

\subsection{Overall Performance}\label{exp_result}

We evaluate our MBGR framework through comprehensive experiments on both generative capability and downstream application performance. All experiments were conducted using the infrastructure described in Section~\ref{Experiments}, with results averaged across three runs to ensure statistical significance. This section presents our key findings, demonstrating the effectiveness of our approach across different business categories, with particular focus on improving performance for smaller businesses.

\subsubsection{Generative Performance}

Table~\ref{tab:generative_performance} presents the Hit@10 performance across different business categories for all models. Our MBGR framework demonstrates significant improvements, particularly for smaller businesses.

\begin{table}[h]
\centering
\setlength{\tabcolsep}{4pt} 
\footnotesize
\caption{Hit@10 Performance Across Business Categories}
\label{tab:generative_performance}
\begin{tabular}{lccccc}
\hline
Model & All & A (61.47\%) & B (9.56\%) & C (12.31\%) & D (16.66\%) \\
\hline
SASRec & 0.0192 & 0.0218 & 0.0101 & 0.0178 & 0.0269 \\
Tiger & 0.0202 & 0.0221 & 0.0128 & 0.0180 & 0.0278 \\
HSTU & 0.0214 & 0.0231 & 0.0127 & 0.0198 & 0.0299 \\
Transformer-MBGR & 0.0341 & 0.0245 & 0.0412 & 0.0321 & 0.0387 \\
HSTU-MBGR & \textbf{0.0410} & \textbf{0.0252} & \textbf{0.0554} & \textbf{0.0398} & \textbf{0.0421} \\
\hline
\end{tabular}
\end{table}

Regarding the Seq-to-Seq architecture, we evaluate two variants of our MBGR framework:
\begin{itemize}
    \item \textbf{Transformer-MBGR}: Uses a standard Transformer encoder-decoder architecture with 16 layers and 8 attention heads, following the architecture described in Section 4.2. The model processes input sequences of maximum length 1500 items and generates Semantic ID token sequences autoregressively.

    \item \textbf{HSTU-MBGR}: Based on the HSTU architecture \cite{hstu}, which introduces a hierarchical transduction mechanism to handle high-cardinality item spaces. We extend it with our proposed BID and MBP modules while preserving its core architecture.
\end{itemize}

Both variants incorporate our proposed BID and MBP modules, with HSTU-MBGR achieving the best performance across all business categories, particularly for smaller businesses (B and C).

\subsubsection{Downstream CTCVR Performance}
Table~\ref{tab:downstream_performance} compares the baseline model with our MBP-enhanced approach.

\begin{table}[h]
\centering
\setlength{\tabcolsep}{3pt} 
\small
\caption{Downstream CTCVR GAUC Comparison}
\label{tab:downstream_performance}
\begin{tabular}{lccccc}
\hline
Model & All & A (61.47\%) & B (9.56\%) & C (12.31\%) & D (16.66\%) \\
\hline
Baseline & 0.7748 & 0.7080 & 0.7594 & 0.8852 & 0.7466 \\
MBGR & \textbf{0.8040} & \textbf{0.7135} & \textbf{0.8258} & \textbf{0.9052} & \textbf{0.7717} \\
\hline
Improvement & +3.8\% & +0.8\% & +8.7\% & +2.3\% & +3.4\% \\
\hline
\end{tabular}
\end{table}

We evaluate the downstream conversion performance using CTCVR GAUC metrics across business categories.

\subsection{BID Encoder Effectiveness}

\begin{figure}[htbp]
\centering
\includegraphics[width=\linewidth]{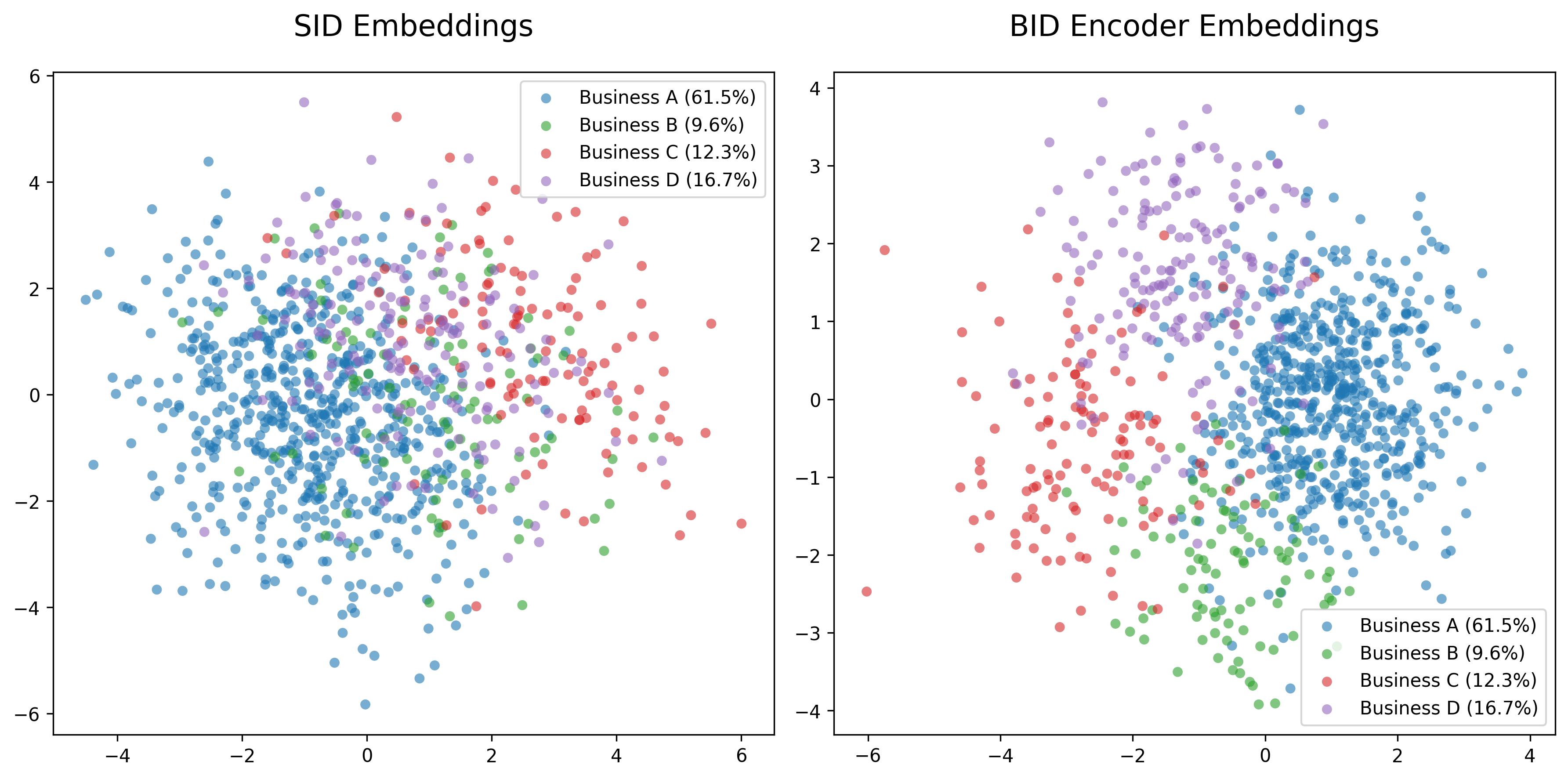}
\caption{Comparison of embedding distributions between sum pooling and BID encoder approaches. The left panel shows embeddings generated by simple sum pooling of token embeddings, while the right panel displays embeddings produced by our BID encoder. All embeddings are projected to 2D space using PCA for visualization. Different colors represent different business types with their proportions: A (61.47\%), B (9.56\%), C (12.31\%), and D (16.66\%).}
\label{fig:embedding_comparison}
\end{figure}

To evaluate the effectiveness of our BID encoder in learning business-aware representations, we conduct a comparative analysis of embedding distributions using Principal Component Analysis (PCA) for dimensionality reduction. Figure~\ref{fig:embedding_comparison} compares the 2D projections of embedding spaces learned by the baseline sum pooling approach and our BID encoder.

The visualization reveals three key insights about the BID encoder's effectiveness:

\noindent 1) \textbf{Business Separation}: The BID encoder (right) shows clearer separation between business types compared to the sum pooling approach (left), particularly for smaller businesses (B and C).

\noindent 2) \textbf{Natural Correlations}: Both methods show natural overlaps between business types, reflecting real-world business correlations. However, the BID encoder maintains these correlations while achieving better separation.

\noindent 3) \textbf{Proportion Alignment}: The clustering patterns align well with the actual business proportions, validating our business-aware design.

These findings demonstrate that the BID encoder successfully captures business-specific patterns while maintaining semantic consistency across different business types.

\subsection{Ablation Study}

We analyze the contributions of each component through ablation studies evaluated on Hit@10 performance. Table~\ref{tab:ablation} shows the impact of removing key components from our full model.

\begin{table}[h]
\centering
\setlength{\tabcolsep}{4pt} 
\footnotesize
\caption{Ablation Study Results on Hit@10 Performance}
\label{tab:ablation}
\begin{tabular}{l|ccccc}
\toprule
Model(HSTU based) & All & A (61.47\%) & B (9.56\%) & C (12.31\%) & D (16.66\%) \\
\midrule
w/o LDR  & 0.0268 & 0.0215 & 0.0312 & 0.0245 & 0.0301 \\
w/o MBP  & 0.0330 & 0.0241 & 0.0401 & 0.0312 & 0.0365 \\
w/o BID  & 0.0335 & 0.0243 & 0.0408 & 0.0318 & 0.0372 \\
Full MBGR & \textbf{0.0410} & \textbf{0.0252} & \textbf{0.0554} & \textbf{0.0398} & \textbf{0.0421} \\
\bottomrule
\end{tabular}
\end{table}
The implementation details for each ablation setting are as follows:
\begin{itemize}
    \item \textbf{w/o LDR}: Instead of business-specific target routing, we use standard next-item prediction that computes loss against all subsequent items regardless of business type. This removes the adaptive target selection mechanism while maintaining single-target decoding.

    \item \textbf{w/o MBP Module}: We simplify the multi-business prediction architecture into a single universal prediction head. This means generating one unified item representation $\mathbf{e}^u$ instead of business-specific representations $\mathbf{e}^b$, and using shared parameters for all business prediction tasks.

    \item \textbf{w/o BID Module}: We replace the business-aware codec with basic semantic transformation:1)Encoder: Use simple mean pooling over token embeddings $\mathbf{t}_i$ instead of the adaptive fusion in BID.2)Decoder: Implement shallow MLP that transforms $\mathbf{e}^b$ to token representations without semantic gate control.
\end{itemize}
 
Key findings:
\begin{itemize}
    \item \textbf{LDR Module} contributes the most significant improvement 
    \item \textbf{MBP Module} provides substantial gains, particularly for smaller business 
    \item \textbf{BID Module} contributes meaningfully across all businesses
    \item All components show synergistic effects, with their combination yielding the best performance
\end{itemize}

\subsection{Hyperparameter Analysis}

We conduct comprehensive experiments to analyze the impact of key hyperparameters, with particular focus on the time decay coefficient $\alpha$ and business weights $w_b$ in the InfoNCE loss function. All experiments are conducted on the generative training dataset with HSTU as the base model.

\subsubsection{Time Decay Coefficient ($\alpha$)}

The time decay coefficient $\alpha$ in $w_t = \exp\left(-\alpha \cdot (t_{\text{last}} - t_{u,t+1}^{(k)})\right)$ controls how rapidly the influence of historical interactions diminishes. We evaluate five settings across different business categories:

\begin{table}[h]
\centering
\caption{Impact of Time Decay Coefficient $\alpha$ on Hit@10}
\label{tab:alpha}
\begin{tabular}{lccccc}
\hline
$\alpha$ & All & A & B & C & D \\
\hline
0.01 & 0.0382 & 0.0241 & 0.0512 & 0.0371 & 0.0403 \\
0.05 & \textbf{0.0410} & \textbf{0.0252} & \textbf{0.0554} & \textbf{0.0398} & \textbf{0.0421} \\
0.10 & 0.0397 & 0.0248 & 0.0531 & 0.0385 & 0.0412 \\
0.20 & 0.0375 & 0.0239 & 0.0498 & 0.0362 & 0.0391 \\
0.50 & 0.0341 & 0.0225 & 0.0442 & 0.0328 & 0.0357 \\
\hline
\end{tabular}
\end{table}

Key observations:
\begin{itemize}
    \item Optimal performance is achieved at $\alpha=0.05$, balancing between recent and historical interactions
    \item Smaller businesses (B, C) benefit more from moderate decay ($\alpha=0.05-0.10$), suggesting their user preferences evolve faster
    \item Larger business (A) shows better tolerance to smaller $\alpha$, indicating more stable user interests
    \item Excessive decay ($\alpha \geq 0.20$) significantly hurts performance across all businesses
\end{itemize}

\subsubsection{Business Weight ($w_b$) Configuration}

The business weight $w_b$ in the InfoNCE loss function is designed to balance the contribution of different business types based on their characteristics. We employ a heuristic weighting strategy derived from empirical analysis of business dynamics:

\begin{table}[h]
\centering
\caption{Business Weight Configuration}
\label{tab:wb_config}
\begin{tabular}{lcccc}
\hline
Business & Proportion & Weight ($w_b$) & Rationale \\
\hline
A & 61.47\% & 0.9 & \makecell{Stable user preferences,\\lower update frequency} \\
B & 9.56\% & 1.5 & \makecell{Fast-changing preferences,\\higher business value} \\
C & 12.31\% & 1.3 & \makecell{Moderate dynamics,\\strategic importance} \\
D & 16.66\% & 1.0 & \makecell{Balanced characteristics,\\baseline weight} \\
\hline
\end{tabular}
\end{table}

The weighting strategy follows these principles:
\begin{itemize}
    \item \textbf{Inverse Relationship to Size}: Smaller businesses receive higher weights to counteract the natural dominance of larger businesses in gradient updates
    
    \item \textbf{Preference Dynamics}: Businesses with faster-changing user preferences (e.g., B) are assigned higher weights to ensure adequate model adaptation
    
    \item \textbf{Business Value}: Strategic business lines receive moderate weight increases to align with platform objectives
    
    \item \textbf{Normalization}: Weights are normalized such that $\sum_b w_b = B$ to maintain overall loss scale
\end{itemize}

Comparative experiments validate this configuration:

\begin{table}[h]
\centering
\caption{Impact of Business Weight Configuration on Hit@10}
\label{tab:wb_results}
\begin{tabular}{lccccc}
\hline
Configuration & All & A & B & C & D \\
\hline
Uniform ($w_b=1.0$) & 0.0368 & 0.0235 & 0.0487 & 0.0342 & 0.0381 \\
Inverse Frequency & 0.0360 & 0.0243 & 0.0521 & 0.0375 & 0.0379 \\
Empirical (Ours) & \textbf{0.0410} & \textbf{0.0252} & \textbf{0.0554} & \textbf{0.0398} & \textbf{0.0421} \\
\hline
\end{tabular}
\end{table}

Key findings:
\begin{itemize}
    \item Our empirical weighting improves overall performance by 11.4\% over uniform weighting
    \item Particularly effective for smaller businesses (B: +13.8\%, C: +16.4\%)
    \item Inverse frequency helps smaller businesses but hurts overall performance
\end{itemize}

\subsubsection{Number of Experts ($K$) in MBP}

We analyze the impact of expert count in the Multi-Business Prediction module:

\begin{table}[h]
\centering
\caption{Impact of Expert Count on Hit@10}
\label{tab:experts}
\begin{tabular}{lccccc}
\hline
$K$ & All & A & B & C & D \\
\hline
4 & 0.0379 & 0.0242 & 0.0503 & 0.0361 & 0.0392 \\
8 & \textbf{0.0410} & \textbf{0.0252} & \textbf{0.0554} & \textbf{0.0398} & \textbf{0.0421} \\
16 & 0.0401 & 0.0249 & 0.0538 & 0.0387 & 0.0413 \\
32 & 0.0387 & 0.0243 & 0.0512 & 0.0372 & 0.0398 \\
\hline
\end{tabular}
\end{table}

Observations:
\begin{itemize}
    \item Optimal at $K=8$ for all businesses
    \item Larger $K$ shows diminishing returns due to parameter fragmentation
    \item Smaller businesses benefit more from moderate expert counts
\end{itemize}

\subsection{Online Experiment Results}

\begin{figure}[hbtp]
\centering
\includegraphics[width=\linewidth, height=1\textheight, keepaspectratio]{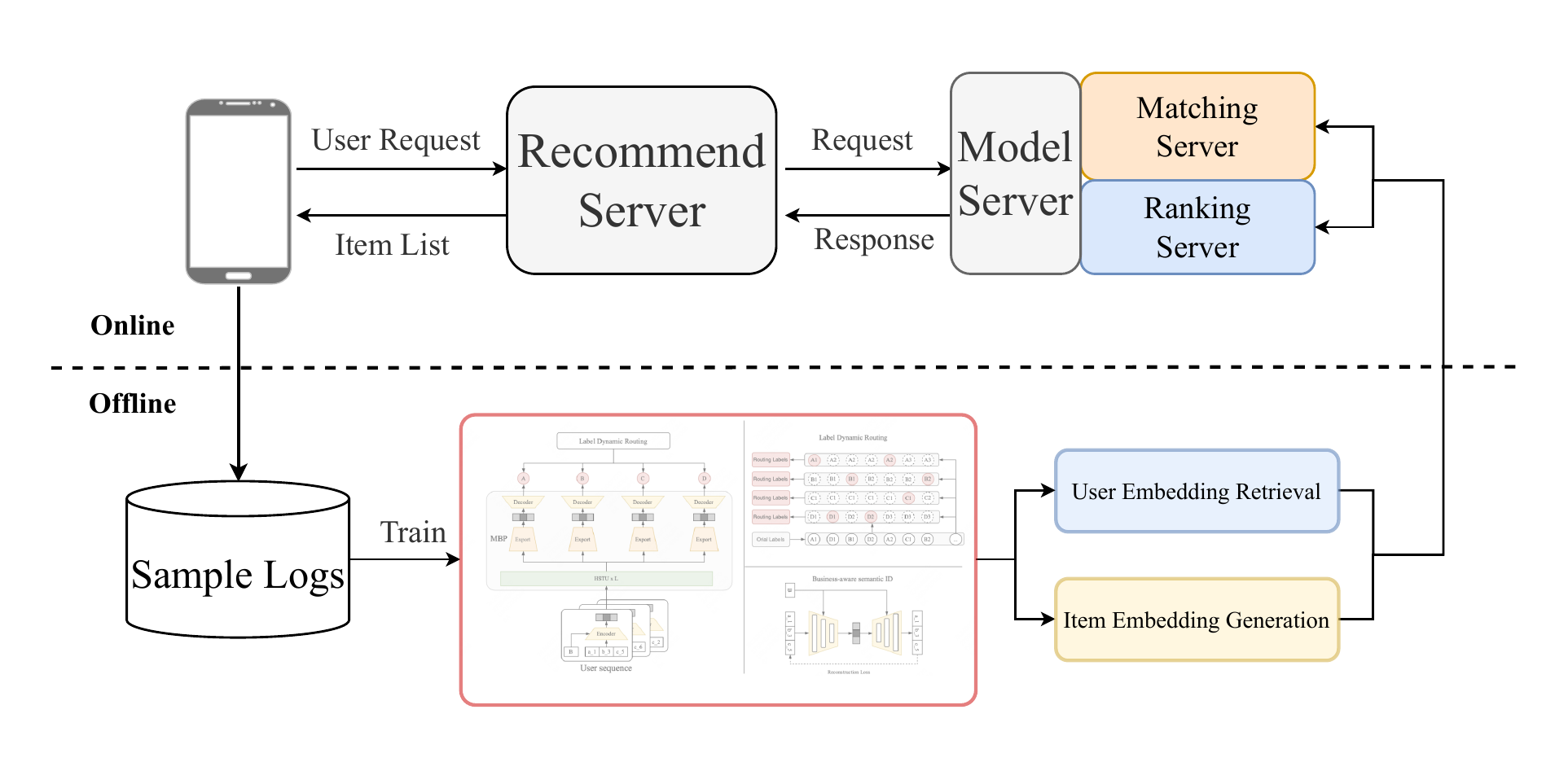}
\caption{Architecture of the online deployment with MBGR.}
\label{fig:mbgr_online}
\end{figure}

As shown in Figure \ref{fig:mbgr_online}, the MBGR framework has been deployed in Meituan's Real-Time Bidding (RTB) advertising system, addressing high queries per second (QPS) and low latency requirements while enabling flexible adaptation to downstream tasks.

In contrast to end-to-end paradigms like OneRec \cite{onerec} and EGA \cite{ega-v2}, which unify the entire recommendation pipeline into a single model, our approach prioritizes industrial robustness. Such unified models entail significant engineering overhead and, more critically, pose substantial instability risks under the highly volatile conditions of Real-Time Bidding (RTB) systems.
Therefore, we designed a more reliable serving strategy for MBGR based on incremental integration. Rather than a full pipeline reconstruction, the generated user and item embeddings are utilized in two ways: as an additional retrieval channel and as enriched features for the ranking model. This architecture ensures system stability while effectively leveraging the representational power of MBGR.
The online serving architecture consists of the following components:

\begin{itemize}[leftmargin=*]
\item{\textbf{User Embedding Retrieval.}} Direct retrieval from distributed cache with precomputed business-specific embeddings.

\item{\textbf{Item Embedding Generation.}} Two-stage process that involves token-level embedding look-up followed by online encoding via the BID encoder.
\end{itemize}

We conducted a one-week online A/B test with 30\% of Meituan's production traffic to evaluate our MBGR framework against the baseline model without MBP enhancement. The experiment was conducted under real-world conditions with live user interactions.

\begin{table}[h]
\centering
\caption{Business-wise CTCVR Improvements}
\label{tab:online_business}
\begin{tabular}{lcc}
\hline
Business & Proportion & CTCVR Improvement \\
\hline
A & 61.47\% & +3.0\% \\
B & 9.56\% & +7.5\% \\
C & 12.31\% & +4.5\% \\
D & 16.66\% & +5.2\% \\
\hline
\textbf{Weighted Average} & 100\% & \textbf{+3.98\%} \\
\hline
\end{tabular}
\end{table}

\textbf{Overall Performance}: Our MBGR framework achieved a statistically significant +3.98\% improvement in CTCVR compared to the baseline (p < 0.01).

The online results demonstrate consistent performance improvements across all business categories, with greater gains for smaller businesses, matching our offline findings. The weighted average improvement of +3.98\% confirms the effectiveness of our approach in production while maintaining robust system performance under high traffic loads.

\section{Conclusion}
In this paper, we identify two key limitations of existing generative recommendation methods in multi-business scenarios: the \textbf{seesaw phenomenon} in NTP-based models and \textbf{representation confusion} in unified SID spaces. To address these challenges, we propose MBGR, a tailored generative recommendation framework for multi-business prediction. MBGR consists of three core components: a Business-aware semantic ID (BID) module that maintains semantic integrity through domain-aware tokenization, a Multi-Business Prediction (MBP) structure that delivers business-specific forecasting, and a Label Dynamic Routing (LDR) module that enhances generation capability by converting sparse multi-business labels into dense representations. Offline and online experiments validate the effectiveness of MBGR in multi-business recommendation, and we have successfully deployed MBGR on Meituan's food delivery platform.
\bibliographystyle{ACM-Reference-Format}
\bibliography{main}

\end{document}